# How Do Shock Waves Define the Space-Time Structure of Gradual Solar Energetic Particle Events?

**Donald V. Reames** (https://orcid.org/0000-0001-9048-822X )
Institute of Physical Science and Technology,
University of Maryland, College Park, MD, USA

**Abstract** We revisit the full variety of observed temporal and spatial distributions of energetic solar protons in "gradual" solar energetic-particle (SEP) events resulting from the spatial variations in the shock waves that accelerate them. Differences in the shock strength at the solar longitude of a spacecraft and at the footpoint of its connecting magnetic field line, curved by solar rotation nominally 55º to the west, drive much of that variation. The shock wave itself, together with energetic particles trapped near it by self-amplified hydromagnetic or Alfvén waves, forms an underlying autonomous structure. This structure can drive across magnetic field lines intact, spreading proton intensities in a widening SEP longitude distribution. During the formation of this fundamental structure, historically called an "energetic storm particle" (ESP) event, many SEPs leak away early, amplifying waves as they flow along well-connected field lines and broaden the distribution outward; behind the structure, between the shock and the Sun, a "reservoir" of quasi-trapped SEPs forms. Very large SEP events are complicated by additional extensive wave growth that can spread an extended ESP-like trapping region around the Sun throughout most of the pre-shock event. Here SEP intensities are bounded at the "streaming limit," a balance between proton streaming, which amplifies waves, and scattering reduces the streaming. The multiplicity of shock-related processes contributing to the observed SEP profiles causes correlations of the events to be poorly represented by the single peak intensity commonly used. In fact, the extensive spatial distributions of SEPs are sometimes free and sometimes interwoven with the structures of the shocks that have accelerated them. We should consider new questions: Which extremes of the shock contribute most to a local SEPs profile of an event, (1) the shock at the longitude of a spacecraft, (2) the shock ~55º to the west at the footpoint of the field, or (3) SEPs that have collected in the reservoir? How does the space-time distribution of SEPs correspond with the underlying space-time distribution of shock strength?

# 1 Introduction

*What are the possible distributions of high-energy protons in large solar energetic-particle (SEP) events, what physical processes produce them, and how do they evolve in space and time?*  With decades of observations on single and multiple spacecraft, aspects of this question still remain unanswered.  Spacecraft observations with adequate energy coverage began to sample SEP spatial distributions in the 1960's and Kahler et al. (1984) established a firm 96% connection between large SEP events and shock waves driven by wide, fast coronal mass ejections (CMEs).  Such a connection had actually been suggested two decades earlier by the type II radio bursts produced by those shocks (Wild et al. 1963).  Yet, it still took time to establish that the shock acceleration of SEPs would supersede the "solar-flare myth" (Gosling 1993, 1994), but that evolution is now well documented in many different ways (Reames 1995, 1999, 2013, 2021a, 2021b; Reames et al. 1997; Zank et al. 2000; Kahler 2001; Cliver et al. 2004; Lee 2005; Cliver and Ling 2007; Rouillard et al. 2011, 2012; Gopalswamy et al. 2012; Lee et al. 2012; Desai and Giacalone 2016; Kouloumvakos et al. 2019).  These "gradual" SEP events contrast with "impulsive" SEP events, with extreme enhancements of $^3$He and heavy elements (Mason 2007) associated with turbulence (e.g. Temerin and Roth 1992) and magnetic reconnection (e.g. Drake et al. 2009), respectively, in solar jets (Bučík 2020).  Impulsive SEP events are usually small and are characterized by their unusual abundances, while gradual SEP events are typically the large "proton events" we discuss here where fast wide CME-driven shock waves generally sample the ions of the solar corona.

Lacking multiple spacecraft conveniently spaced in solar longitude during single SEP events, Cane et al. (1988) sorted the intensity-time profiles of 235 different events observed by the *Interplanetary Monitoring Platform* (IMP) 4, 5, 7, and 8, and the *International Sun-Earth Explorer* (ISEE 3) spacecraft as functions of their solar-source longitude.  Owing to the Parker spiral of the solar magnetic field produced by solar rotation, spacecraft near the Earth are magnetically connected to a nominal longitude of W55 on the Sun (for a nominal solar wind speed of 400 km s$^{-1}$).  A fast shock source centered on W55 could send its strongest burst of particles out the field line early in an event with profiles reasonably explained by energy-dependent pitch-angle scattering (Parker 1963; Ng et al. 1999, 2003; Desai and Giacalone 2016); as this shock expands radially, the Parker spiral causes this magnetic connection point to sweep eastward along the eastern flank of the shock, most likely along a weakening source.  Sources farther to our west than W55 would begin on the eastern flank of the shock, presumably weaker than the shock nose, and would become weaker still as the connection point moved farther east and the shock moves outward in radius. Thus, western solar sources are initially strong but fade rapidly.  In contrast, SEP sources to our east begin on the weaker western flank of the shock but our connection point swings eastward with time; such an improving connection toward the shock nose can somewhat offset the effect of the radially diminishing shock strength.  Thus, spacecraft on the western flank of the shock (eastern sources) can observe stronger effects late in the event, near the time of local shock passage. This overall structure has been rather poorly documented although it has been initially explored theoretically, for example, by the extensive effort of Lee (2005).

Multi-spacecraft observations were studied productively by McKibben (1972), for example, by combining IMP 4 observations with those on *Pioneer* 6 and *7*, all at 1 AU, but the era of the *Helios* spacecraft, together with IMP 8 provided a unique moderately-spaced cluster of spacecraft (e.g. Lario et al. 2006), especially when augmented by the *Voyager* spacecraft that provided a 2 – 4 AU backup (Reames et al. 1996, 1997, 2012). We will revisit and extend the coverage of some of these observations below.

The *Solar Terrestrial Relations Observatory* (STEREO) *Ahead* (A) and *Behind* (B) spacecraft were launched in 2006 at the end of Solar Cycle 23 and were widely separated, each >90⁰ from Earth, when events of the new solar cycle began to occur in 2010 and 2011. The STEREO events have been extensively analyzed, usually by fitting the peak intensities of electrons or protons of specific energies at each of three spacecraft to a Gaussian distribution (e.g. Lario et al. 2013; Paassilta et al. 2018 and references therein). Unfortunately, peak intensities at different longitudes frequently occur at different times, mixing variations in space and time; they involve different regions of the shock and even different physical mechanisms, so these fits can vary widely.

Cane et al. (1988) showed typical intensity-time profiles of SEP events and sketched the way these distributions typically varied with solar source longitude, and subsequent review articles (e.g. Reames 1999, 2013, 2021a) duplicated these distributions using actual proton observations from different solar longitudes at several energies. A version of this figure is shown as Fig. 1.

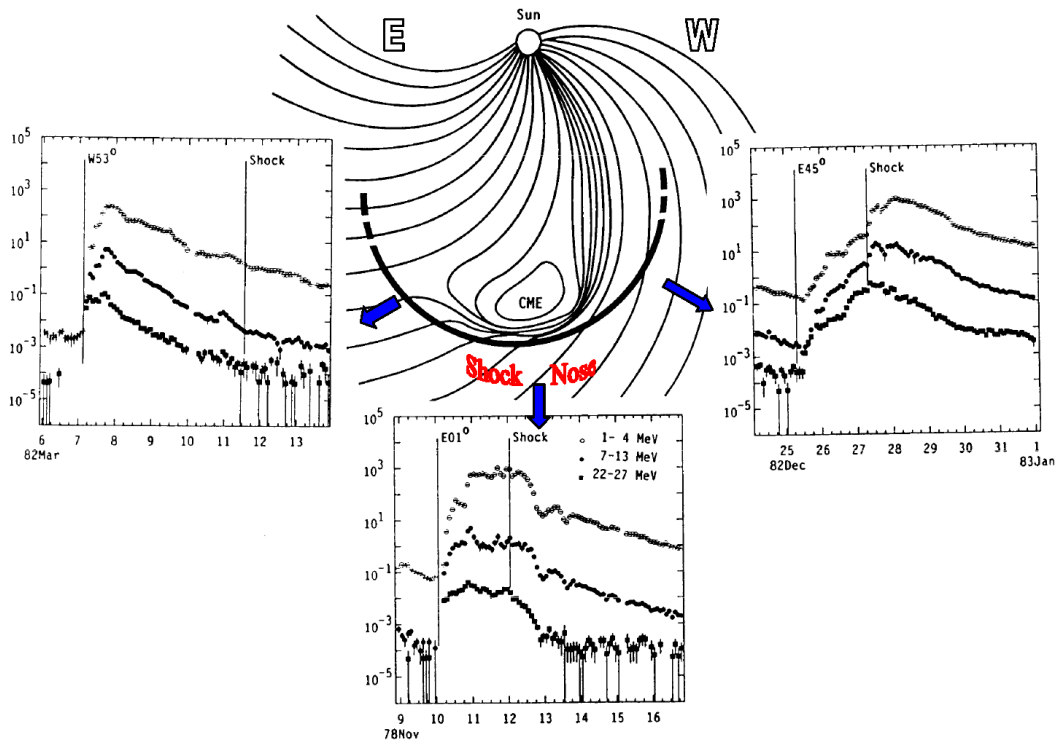

**Fig. 1** Typical intensity vs. time profiles of SEPs as a function of solar longitude presented in review articles (e.g. Reames 1999, 2013, 2021a) since Cane et al. (1988). These profiles show samples of the dominant effects on SEPs of the longitudinally varying strength of the CME-driven shock wave and of the Parker spiral magnetic field caused by solar rotation (see text). This field curvature causes an observer's magnetic footpoint longitude to lie ~55° further to its west.

The event on the east flank of the shock in Fig. 1 is well connected to the strongest "nose" of the shock early and receives the most intense burst of particles, but is subsequently connected to the weaker and weaker shock flanks farther out from the Sun. The event near central meridian is connected to the weaker west flank early but receives a strong impetus from the shock nose when it passes the spacecraft despite its diminishing strength as it has moved out radially. The central observer may also see a SEP depression during passage through the driving CME. On the western flank in Fig. 1, an observer is connected to a very weak-but-strengthening shock region early, but may receive maximum intensity after crossing the local shock and arriving on field line that connect to the stronger shock nose from behind.

The profiles in Fig.1 do illustrate the effects of shock variation and magnetic curvature that are typical. However, typical SEP events are not the whole story, and some atypical events, which have been ignored for decades, may tell us more about the physics of SEP acceleration. Unique features seen in early multi-spacecraft studies (Reames et al. 1996, 1997) suggest broader consequences that invite further study. The present article reviews and expands observational coverage of SEP distributions of protons of ~2 − 200 MeV and suggests connections between these distributions and the properties and evolution of the shock waves that accelerate them. We investigate the possibility that SEPs, trapped in spatial structures composed of self-amplified Alfvén waves, are a stable underlying factor revealing spatial evolution.

Data of the SEP observations discussed in this article are generally available at https://cdaweb.gsfc.nasa.gov/.  Data on CME speeds were obtained from the *Large Angle and Spectrometric Coronagraph* (LASCO) on the *Solar and Heliospheric Observatory* (SOHO) at https://cdaw.gsfc.nasa.gov/CME_list/.  Sources of other observations are listed as they are discussed.

## 2 A Great Variety of SEP Intensity-Time Profiles

If distributions of all SEP events were the same, they could be mapped by one spacecraft observing many events from differing solar longitudes.  Even though events differ, such observations can show similarities that depend upon the global structure of the solar field as well as differences dependent upon features of the shock wave, the SEP event, and local field.  Figure 2 shows time profiles of protons of broad multiple energies for seven SEP events observed by IMP 8 with varying source longitudes distributed around a central map showing their distribution relative to a fixed CME/shock source driving downward to provide shock-relative coordinates, as in Fig. 1.  When available, we have added at the event onset time, the source longitude and the CME speed from SOHO/LASCO.  For events where the local shocks have been analyzed, the time of shock passage of the spacecraft is also noted with the shock speed and the angle between the magnetic field vector $\boldsymbol{B}$ and the shock normal, $\theta_{Bn}$ from analysis of *Wind* data by J. Kasper (http://www.cfa.harvard.edu/shocks/wi_data/).  Of course there is no single CME or shock speed and span for all events, and $\theta_{Bn}$ describes only one point in time and space at the shock, but the available parameters help scale the variations we see.

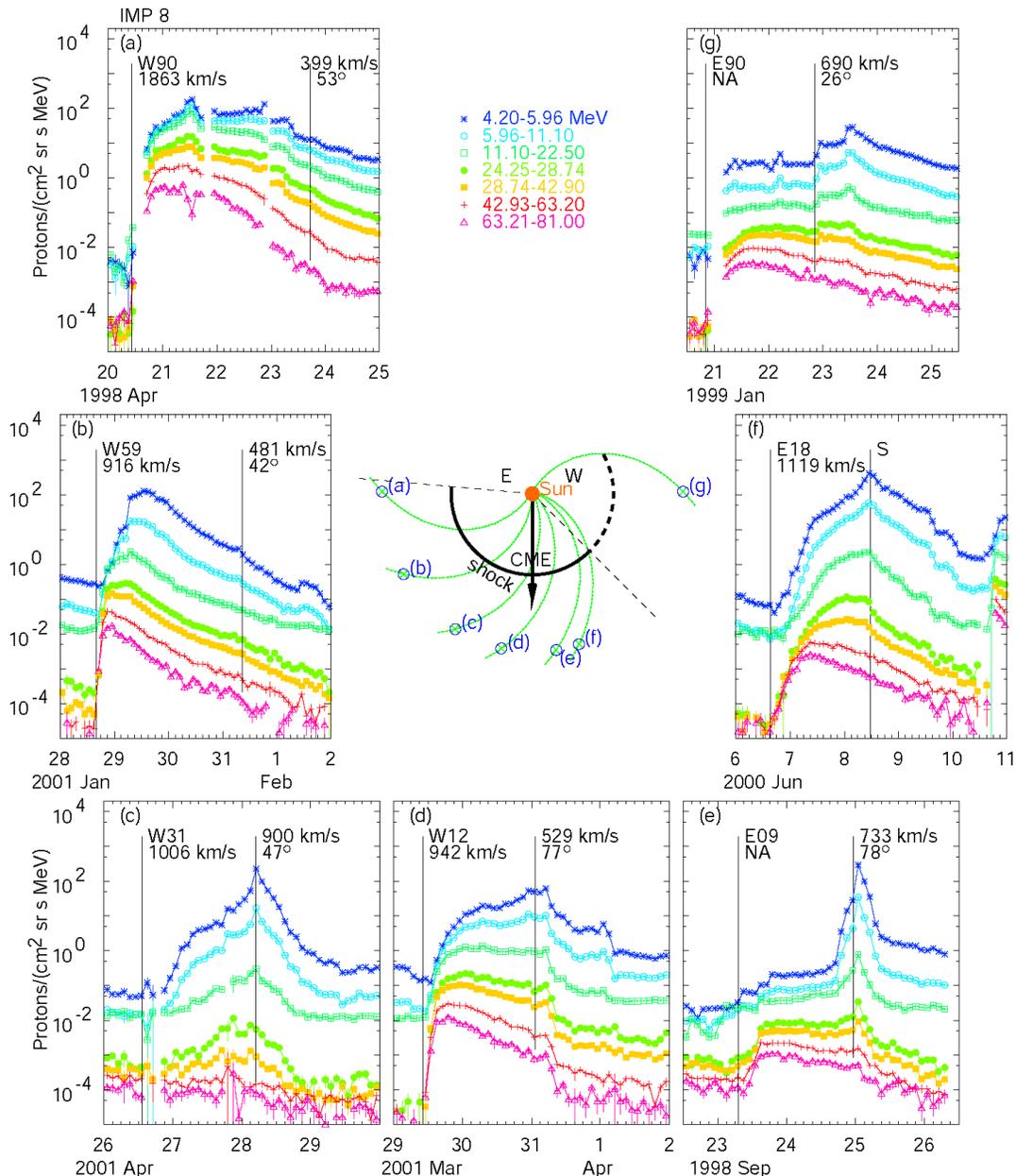

**Fig. 2** IMP 8 proton intensities of the listed energies are shown vs. time in each panel (a) through (f) around a central map showing their nominal distribution around a fixed CME-driven shock source. Event onset times in each panel are flagged with the source longitude and the CME speed (when available) and the times of shock passage are noted with the shock speed and $\theta_{Bn}$ (when available). All panels have the same intensity scale. Dashed lines in the central map illustrate that a shock must first encounter any field line at its footpoint on the east flank, but may first strike it far from the Sun on the west.

As noted in the introduction, the observer's magnetic footpoint longitude typically lies ~55º to the west of the longitude of the spacecraft itself. At which longitude is the shock strongest? On the eastern flank of the shock, this footpoint longitude may reach 55º closer to the shock nose, making it much more important than the local shock, since by contrast, the effect of the latter is diminished both by being at a more-remote longitude from the shock nose and at increased radial distance. On the western flank of the shock, the local shock can be much more significant, being closer to the shock nose than the footpoint is. Thus, the local shock speed is much reduced in Figs. 2a and 2b and has little or no effect on the SEPs.  In Fig. 2c, the CME and local shock speeds are nearly the same, yet the local shock dominates.  Why does the 1006 km s$^{-1}$ CME have so little effect on the early SEPs in Fig. 2c?  Perhaps the early magnetic connection to the CME that we expect has been disrupted.   Often the local shock dominates energies below ~50 MeV on the western flank of the shock (eastern sources) even out at 1 AU.

The peak at the shock like that seen in Fig. 2e was historically called an "energetic-storm-particle" or ESP event, and was often treated as if it was unrelated to SEPs.  The shock peak in this event differs from that in Fig. 2c in having an efficient quasi-perpendicular shock wave with $\theta_{Bn}$=78º vs. only 47º.   It is quite likely that the event in Fig. 2d with $\theta_{Bn}$=77º also has a well-formed ESP event that has been "buried" in the addition of residual earlier SEP emission, beginning with the shock at the footpoint of the field.   The event in Fig. 2e has minimal early injection to compete with the ESPs surrounding the local shock; early protons have leaked away unobserved on different field lines; we will return to this event in Sect. 4.  The dashed line on the central map in Fig 2 near the field lines to events (d) and (e) shows that a weak region on a shock (dashed) can intercept the inner field lines initially while a strong shock (solid) later intercepts the outer field lines leading to an observer.

The early behavior of the SEP profiles can be controlled by the shock in the radial field region at the footpoint of the observer's field line, unbiased by east or west longitude.  The shape of the early rise in the profiles in Fig. 2a and Fig. 2g are similar, except that the former is more intense because it has a stronger footpoint shock. Incidentally, the data gap at the beginning of each of these two events occurs because radio emission from the event disrupts IMP 8 telemetry.  The peak in Fig. 2g comes after the shock, probably on field lines that connect to a stronger region of the shock to the east.  We will discuss other reasons for SEP peaks behind shocks in Sect 3.3.

Figs. 2b, 2d, and 2g resemble the "typical" events identified by Cane et al. (1988), as does the IMP 8 profile discussed in Sect. 3.3, and these properties were sustained in the reviews by Reames (1999, 2013, 2021a) and shown in Fig. 1.  However, events with peaks at the local shock are also common and they carry direct evidence of important connections on underlying physics and structure, as we will see.

# 3 Multi-Spacecraft Distributions

## 3.1 Evolution of Shock Peaks and the ESP Event

Clusters of spacecraft can be greatly helpful in mapping the evolution of SEP spatial distributions, and the period in 1978 when the two *Voyager* spacecraft, somewhat beyond 1 AU, joined by the two *Helios* spacecraft inside 1 AU with IMP 8 near Earth, produced some especially interesting SEP studies like the 1 January 1978 event shown in Fig. 3 that was recognized by Reames et al. (2012).

Figure 3b shows a shock of limited extent originally centered near the field line connected to *Helios 1* that produces a strong early intensity increase in SEPs (Fig. 3a) at that spacecraft which remains high until shock passage.  *Helios 2* and IMP 8 are nearly centrally located and see a shock transit speed of 880 km s$^{-1}$.  At these two spacecraft, proton intensities above ~30 MeV leak out early and peak during the first day, while intensities of protons of lower energies continue to rise up to peak between the times of the shock and a magnetic cloud at each spacecraft (Fig. 3a or Figs. 3c and 3f).

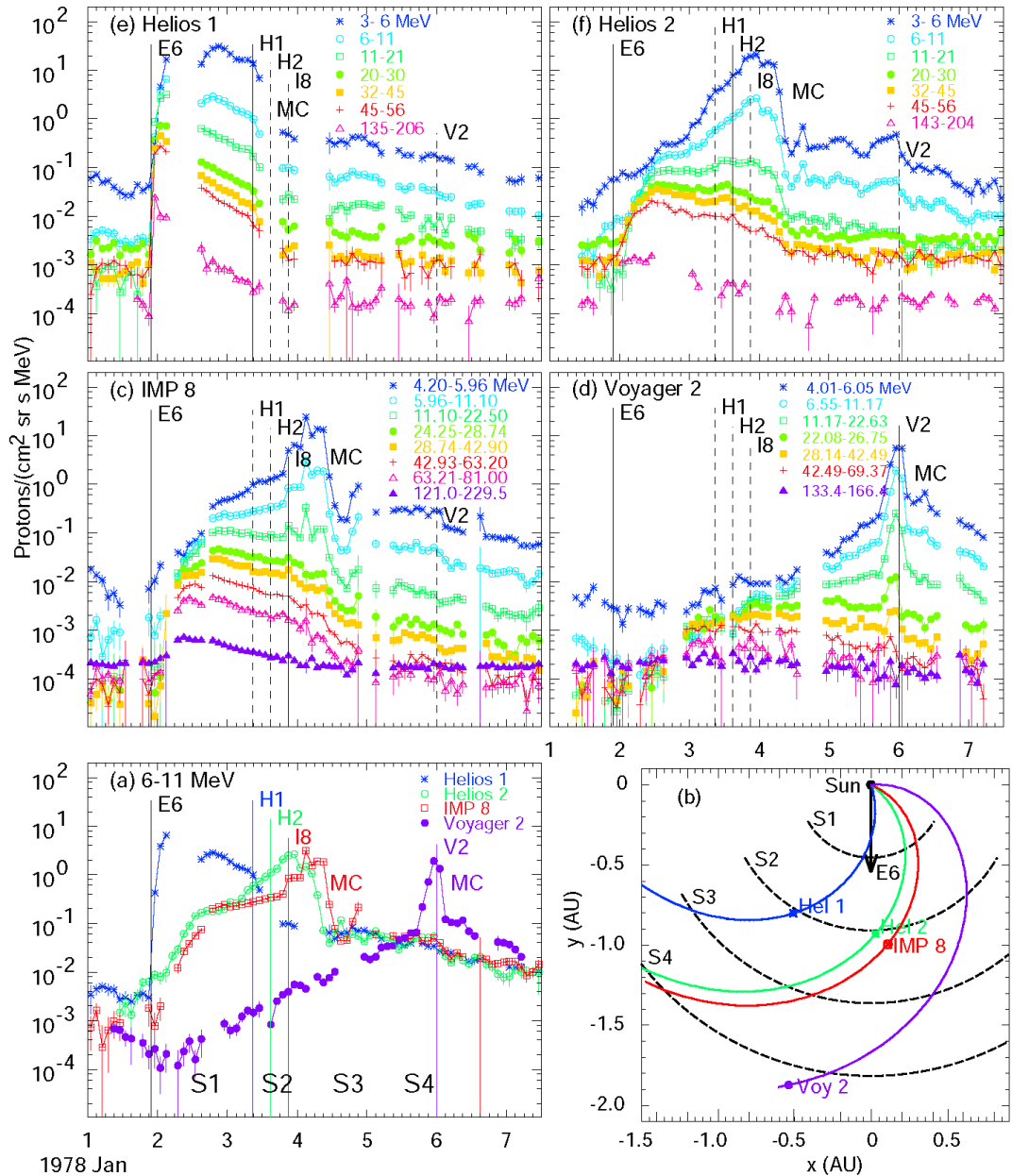

**Fig. 3** In (**a**), intensities of 6 – 10 MeV protons are compared for *Helios 1, Helios 2,* IMP 8, and *Voyager 2* during the 1 January 1978 SEP event, while (**b**) shows the spatial configuration of the spacecraft on their initial field lines and stages in the expansion of the CME-driven shock at S1, S2, etc. are sketched. Intensity-time profiles for a full list of energy intervals are shown and shock arrival times at each spacecraft are shown for (**c**) IMP 8, and (**d**) *Voyager 2,* (**e**) *Helios 1,* and (**f**) *Helios 2.* MC is a magnetic cloud from the original CME. Onset time of the event is flagged by E6 and shock passage at each spacecraft by H1, H2, I8, and V2.

The behavior of *Voyager 2* in Fig. 3a is most interesting; SEP intensities at *Voyager* do not increase until after S1, indicating that the shock does not intercept the field line to *Voyager* (see Fig 3b) before 3 January. Intensities then rise continuously until they suddenly increase in the ESP event as the shock, with its trapped waves and particles, passes *Voyager* followed by a magnetic cloud (Burlaga et al. 1981). In Fig. 3d we see that the higher energies do rise and peak earlier, but evidence of the ESP peak at the shock is still seen up to ~50 MeV. Figure 3d is quite similar to Fig. 2e except for its initial onset delay and slower evolution.

Initially, the extremely well-connected *Helios 1* spacecraft receives an intense burst of protons at energies up to ~200 MeV or more. These outward streaming protons amplify Alfvén upstream waves that scatter subsequent particles, trapping them near the shock. When the shock reaches radius near S1, *Helios 2*, and IMP 8 have begun to see significant outbound proton intensities from the western flank of the shock, but *Voyager 2* remains quiet since the shock does not yet intercept its field line. The SEP peaks at *Helios 2*, and IMP 8 are compressed between the shock and a following magnetic cloud (Burlaga et al. 1981). At S2, all three inner spacecraft see high intensities or peaks near the time of local shock passage at energies below about ~30 MeV, and *Voyager* begins to see quite energetic protons leaking from the flank of the shock which has intercepted its field line. At S3, the inner spacecraft all reside in a "reservoir" (see Sect. 3.3) behind the shock, all spacecraft with nearly the same intensities at all energies, and the intensities at *Voyager* approach those levels. At S4, the ESP peak of particles still captured at the shock arrives at *Voyager* with intensities of protons up to ~50 MeV comparable with those in the earlier shock peaks at the other spacecraft.

This event is an excellent example of the evolution and propagation of a detached peak of SEPs at shock structures we call ESP events. The absence of protons that leaked out early to form this structure highlights the independent existence of the particles trapped at the shock peak that is often buried and obscured. Here, the protons leaked in one place and the structure moved cross-field to another as a naked ESP event with the preceding intensities stripped away.

### 3.2 Extensive Shocks: Double Jeopardy

When strong shock waves subtend a large angle around the Sun, multiple encounters of shock-accelerated SEPs with the field line connected to Earth, for example, become possible, as seen in the SEP event of 23 September 1978 shown in Fig. 4.

At the onset of this event, SEP intensities at well-connected IMP 8 (Fig. 4a and 4c), near Earth, rise very rapidly, while those at *Helios 1*, and *Helios 2*, connected to the western flank of this extensive shock rise more slowly (Fig. 4a, despite data gaps). After arrival of this shock, with transit speed 910 km s$^{-1}$, IMP 8, *Helios 1*, and *Helios 2* join a reservoir where intensities decline as this volume of protons, magnetically trapped behind the shock, expands adiabatically (Fig 4a after S1). Meanwhile, intensities at the *Voyager* spacecraft still slowly increase until S2, where the *Voyagers* and IMP 8 have similar proton intensities on their similar field lines, after the east flank of the shock has moved off of these field lines. *Helios 1* and *Helios 2* have higher intensities at S2 since both are still on field lines with protons trapped behind the shock.

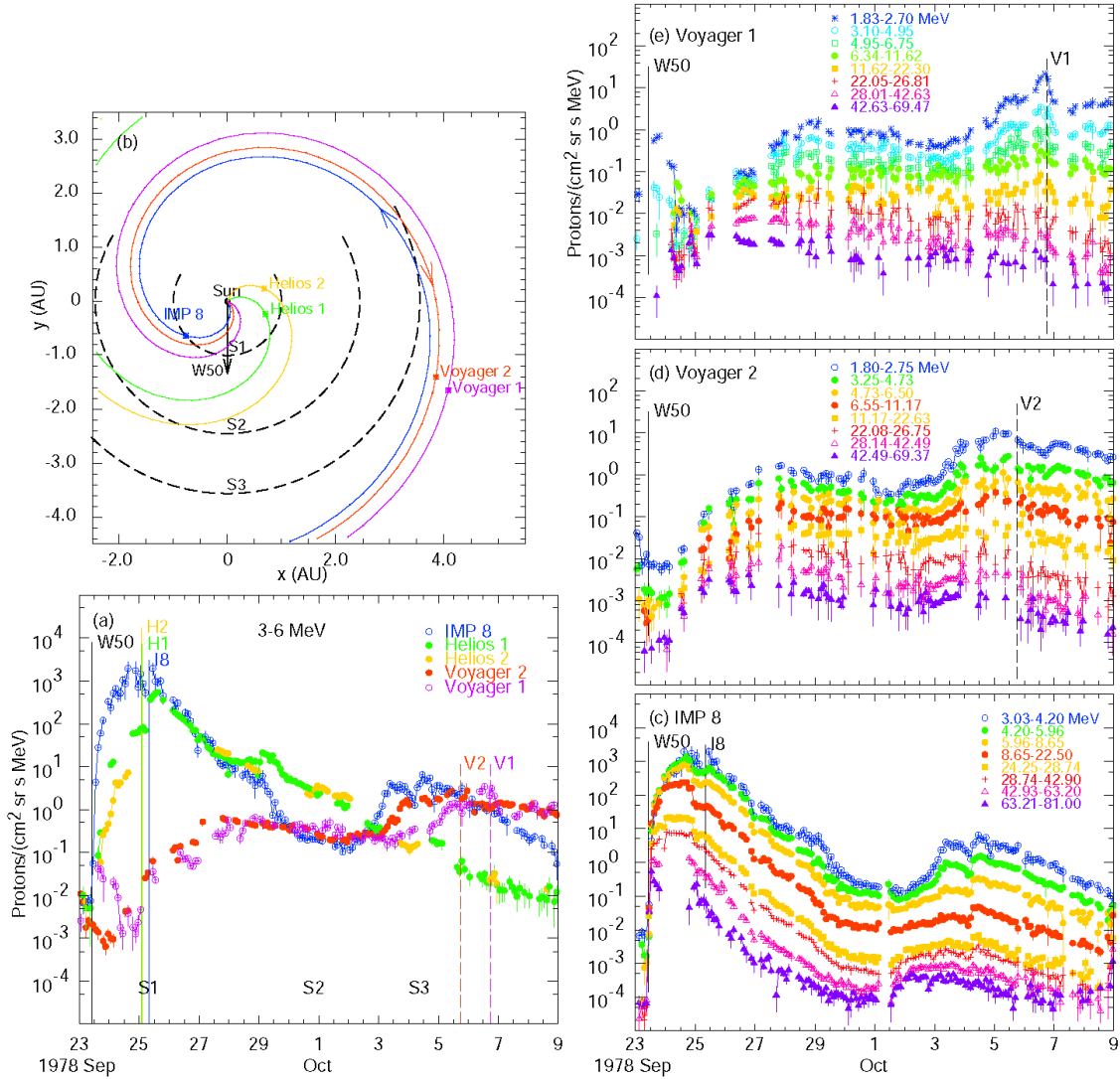

**Fig. 4** In (**a**) intensities of 3 − 6 MeV protons are compared for IMP 8 (blue), *Helios1* (green), *Helios 2* (yellow), *Voyager 1* (red), and *Voyager 2* (violet) in the 23 September 1978 SEP event, while (**b**) shows the spatial configuration of the spacecraft on their initial field lines and stages in the expansion of a CME-driven shock at S1, S2, and S3 are sketched. Onset time of the event is flagged by W50 and shock passage at each spacecraft by H1, H2, I8, V1, and V2. In (**b**), the western flank of the shock S3 intercepts the blue and red fields, where arrows direct particles accelerated sunward to IMP8, then outward to *Voyager 2*, respectively, and to *Voyager 1* later as a residual ESP event. More-complete proton intensities at listed energies are shown for (**c**) IMP 8, (**d**) *Voyager 2*, and (**e**) *Voyager 1*.

At the time of S3 in Fig. 4a intensities of IMP 8, then *Voyager 2*, then *Voyager 1* show new consecutive increases as the *western* flank of the same shock strikes the field lines to each of those three spacecraft as shown in Fig. 4b. Particles are then sent sunward to IMP 8 and outward to *Voyager 2*. In the initial study of this event, Reames et al. (1996) showed IMP 8 proton anisotropies with outward flows early on 23 – 24 September and then sunward flows on the second peak on 3 – 5 October. Fig. 4c shows that high energies up to ~80 MeV are still involved in these late increases, with the highest energies arriving back at IMP 8 first; thus showing velocity dispersion at IMP 8 (Fig. 4c) that is weak at nearby *Voyager 2* (Fig. 4d). Meanwhile, *Voyager 1* (Fig 2e) shows clear evidence of an ESP event on 6 October, peaking outside 4 AU, two weeks after the onset of the original SEP event.   This ESP event is being observed quite far west of the nose of the shock.  The geometry in Fig. 4b strongly suggests that the shock is likely to be quasi-perpendicular at *Voyager*, but differences between the two spacecraft intensities near the shock Figs. 4d and 4e suggest significant spatial variations.

Actually, we are unable to measure the shock arrival at *Voyagers 1* and *2*, (shown dashed) since both arrive during plasma and field data gaps, after which the solar-wind density, speed, and magnetic field intensity have increased significantly.

The reader should realize that the spacecraft configuration with IMP 8, *Helios 1* and *2* and with *Voyager 1* and *2* on similar field lines at ~4 AU is extremely rare. However, out where the field lines have made a complete turn about the Sun, multiple shock strikes must become increasingly common, even inevitable, when the shocks are sufficient strong, but we have few chances to observe them.

### 3.3 Reservoirs

Reservoirs, named by Roelof et al. (1992), have been reviewed before (e.g. Reames 2013), but they are especially relevant here, in a discussion of spatial distributions, because they are invariant in space and time (Reames et al. 1997) and are often described as a magnetically-trapped or quasi-trapped volume of particles behind the expanding shock, populated by particles swept downstream as the shock advances; overall intensities slowly decrease as the effective volume expands adiabatically.

Figure 5 shows a classic event, first discussed by Reames et al. (1996, 1997), where *Helios 1* (green) is well connected near the shock nose early, with rapidly increasing intensities to peak near the time of shock passage.  Intensities at *Helios 2* (orange), then IMP 8 farther west (violet), slowly increase until they enter the reservoir behind their local shock.  At time noted as R, all three spacecraft see the same reservoir energy spectrum, shown on the right in Fig. 5.

Time profiles like that of IMP 8 in Fig. 5, with a gradual rise to a peak, well behind the shock, were correctly considered as typical by Cane et al. (1988) for observers on the western flank of the shock.   In this event, we expect that the shock initially accelerates SEPs with decreasing strength from the nose, near *Helios 1*, around the western flank, past IMP 8, but by the time the shock reaches 1 AU it has become too weak to accelerate particles except near its nose.  Thus, IMP 8 sees increasing intensities from the earlier acceleration as its field line scans eastward along the expanding shock, with nothing extra seen at the now-defunct local shock; intensities only reach a peak later when it enters the well-confined reservoir which then continues its own expansion and intensity decrease.

Profiles like those of *Helios 2* and IMP 8 are common on the western flanks of shocks.  In fact their slow intensity rise may have been the root of the term "gradual." Their magnetic connection scanned to the east among SEPs left behind along a shock whose strength initially increased to the east; nevertheless the ability of that shock to accelerate ions expired before it reached 1 AU.  Their peak intensities are completely unrelated to the point on the shock, if any, that is initially at their magnetic footpoint. These can be described as reservoir-dominated SEP profiles.  When we can see the full spatial structure, the cause of the profile of any single spacecraft, like IMP 8 here, becomes clearer.

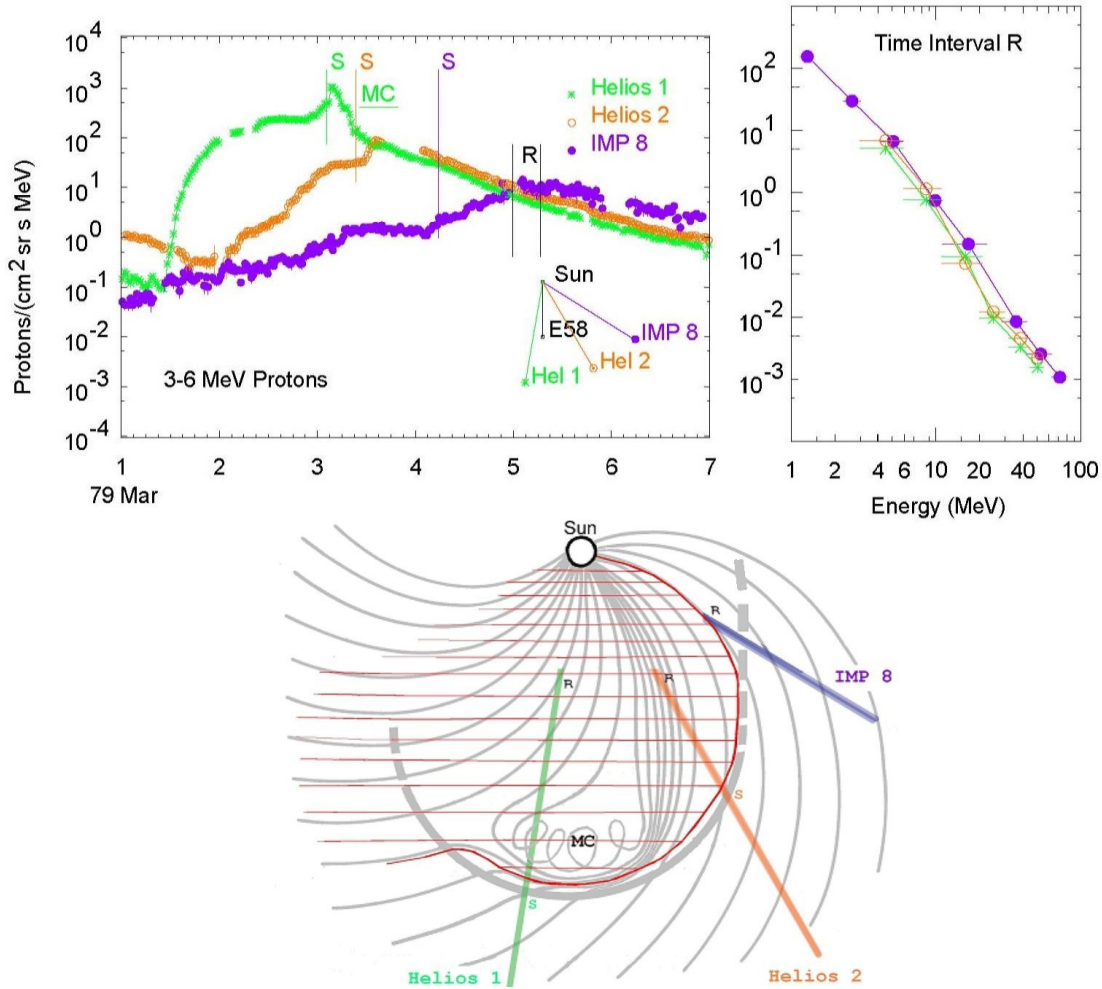

**Fig. 5** The left panel compares intensities of 3 – 6 MeV protons vs. time on *Helios1* (green), *Helios 2* (orange) and IMP 8 (violet) in the event of 1 March 1979. Time of shock passage at each spacecraft is labeled as S in the appropriate color. The right panel compares the full energy spectra on all three spacecraft at point R when they have all entered the expanding and declining reservoir. The lower cartoon shows how the three spacecraft would enter the reservoir region (shaded red) behind the shock and magnetic cloud (MC) which expand past them (actually, of course, the spacecraft are fixed in space during the event and the CME and shock expand out past them).

In Fig. 5 the shock has been extended farther to the west than the east mainly because we know its effects on the west. However, this asymmetric situation is not uncommon in modeled shocks (see Fig. 7 in Rouillard et al. 2011).

### 3.4 STEREO and Gaussian Fits to the Spatial Distribution

As stated above, STEREO A and B were both widely separated from Earth when their early years during solar minimum finally ended. Launched in December 2006 they were ±120° from Earth in September 2012 forming three equally-spaced observation points.

Nearly all of the many spatial studies of STEREO data are based upon fitting the observed peak proton intensities $j_{max}$ at various energies to ad hoc Gaussian distributions of the form $j_{max} = A\ exp\{(\phi - \phi_0)^2\ /\ 2\ \sigma^2\}$, where $\phi$ is the longitude and $\phi_0$, $A$, and $\sigma$ are constants, and $\sigma$ measures the angular half-width of the distribution. This form had also been extensively applied to *Helios* and IMP data by Lario et al. (2006). Xie et al. (2019) studied 19 – 30 MeV protons in 28 events and found $\sigma = 39^o \pm 6.8^o$. Paassilta et al. (2018) compiled a catalog of 46 wide-longitude events above 55 MeV, of which seven were suitable, and found $\sigma = 43.6^o \pm 8.3^o$.

Dresing et al. (2018) made the next step of fitting Gaussian dependence of electrons as a full function of time during an event. Figure 6a shows simple Gaussian fits to the hourly intensities of ~20 MeV protons at *Wind* and STEREO A and B, and Kahler et al. (2023) have extended this study fitting hourly ~20 MeV proton intensities from STEREO and *Wind* in three events, and showing the evolution in Carrington longitudes. As a summary of the time dependence of the 23 January 2012 event, Kahler et al. (2023) show three time slices, two days apart, with the log of the Gaussian amplitude as radius and the angles at half maximum determining the angular borders and width as shown in Fig. 6b.

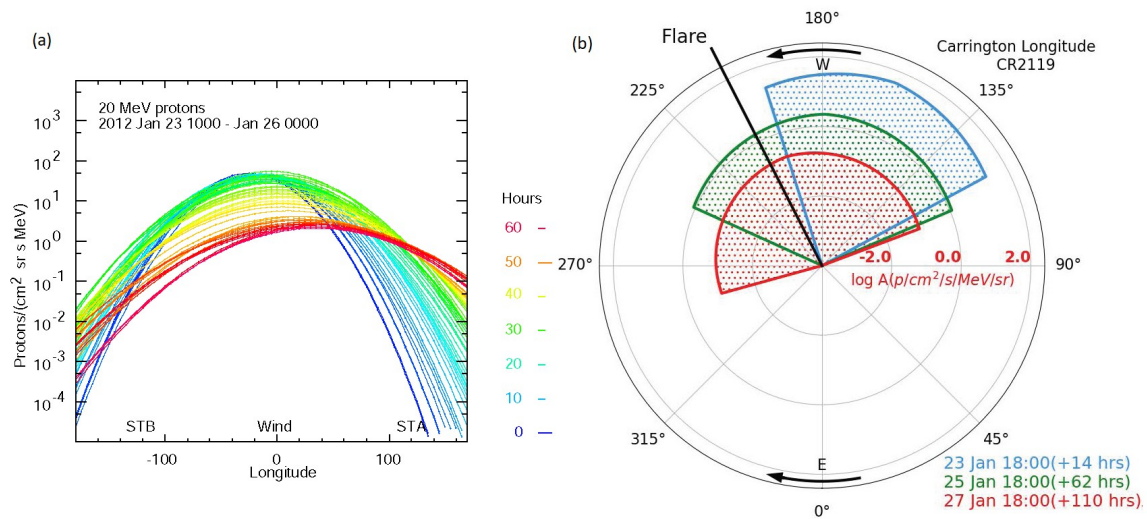

**Fig. 6 (a)** Shows simple hourly Gaussian (parabolic) fits to the intensities of 20 MeV protons at STEREO A , *Wind*, and STEREO B vs. solar longitude as a function of time in the 23 January 2012 SEP event, while **(b)** shows an azimuthal plot of longitudes and amplitudes of modeled Gaussian hourly SEP distributions in Carrington longitude (CL) coordinates at selected times listed (from Kahler et al. 2023). Cone angle widths of each distribution are half intensity points = 2.354 $\sigma$. Initially spacecraft locations are *Stereo A*, *Wind*, and *Stereo B* at CL072, CL186, and CL294, respectively.

The source flare is $50^o$ west of the initial 1 AU distribution centroid in Fig. 6b, as expected from the Parker spiral field. The distribution then spreads westward, increasing its width substantially as the intensity decreases. Spreading in width as a function of time is a common feature of the events, especially westward spreading as the shock waves cross the Parker spiral. Shocks can spread in either direction, but even for radial propagation, the direction of solar rotation causes the bias for westward spread.

# 4 Shock Acceleration and its Spatial Wave Context

The relevant theory of diffusive shock acceleration has been well reviewed (Jones and Ellison 1991; Lee et al. 2012; Desai and Giacalone 2016). Protons streaming away from a shock can generate or amplify Alfvén waves (Stix 1992) that scatter subsequent particles back and forth across the shock with a gain in velocity from Lorentz transformations from the wave frames upstream and downstream on each transit. Particles of magnetic rigidity $P$ and pitch angle cosine $\mu$ resonantly scatter on waves of wave number $k \approx B/\mu P$ in magnetic field intensity $B$. A self-consistent equilibrium theory of particles and self-amplified waves was developed by Bell (1978a, b) and was applied to interplanetary shocks by Lee (1983, 2005). Lee (1983) simplified the behavior by letting $\mu \approx 1$ so that each particle rigidity or energy had its own corresponding resonant wave number. The predicted relationship between particle and wave spectra at interplanetary shocks has been generally confirmed experimentally (e.g. Tsurutani et al. 1983; Viñas et al. 1984; Kennel et al. 1986; Tan et al., 1989). Equilibrium solutions provide spectra and spatial distributions but cannot determine growth rates or the time evolution of the spatial structures. Ng and Reames (2008) supplied a fully self-consistent time-dependent study of the early phase of the balance between wave growth and particle scattering during shock acceleration.

   Figure 7 is a sketch for visualizing and discussing shock and ESP evolution in energy and space. As protons stream away from the shock at $E_1$, they generate or amplify resonant waves that scatter and trap subsequent $E_1$ protons so they scatter across the shock gaining energy on each transit, soon to arrive at $E_2$. At $E_2$ they must again stream out to generate or amplify resonant waves sufficient to scatter and trap $E_2$ protons until they arrive at $E_3$, etc. The acceleration of protons to >300 MeV near the Sun might take ~10 min as shown by the self-consistent, time-dependent calculations of Ng and Reames (2008), but limitations of the turbulence to $\delta B \ll B$ in that simulation make this time scale conservative. Generally, there is a flow of protons from the seed particles injected at the base of the energy distribution and there is proton leakage at each energy as they flow upward; where the upward flow runs out there is a spectral break. Simulations (Ng and Reames 2008) show that when pitch angles are considered, the first particles to arrive at each new energy have $\mu \ll 1$; a proton with rigidity $P$ and $\mu = 0.1$, for example, resonates with waves produced by protons with rigidity $0.2P$ and $\mu = 0.5$, for example, which are much more abundant.

**Fig. 7** A simplified sketch for discussing the evolution of wave growth and SEP energy increases at a shock. As protons stream out from a shock at one energy, they generate waves that trap and scatter subsequent protons back and forth across the shock to gain a higher energy, etc. Where the flow of particles up the energy axis runs out, a spectral break occurs. This classic discussion of shock acceleration, a shock with a field of Alfvén waves with trapped particles, also exactly describes an ESP event. The spatial width of an ESP event defines the extent of the self-amplified wave field, and also depends upon $\theta_{Bn}$.

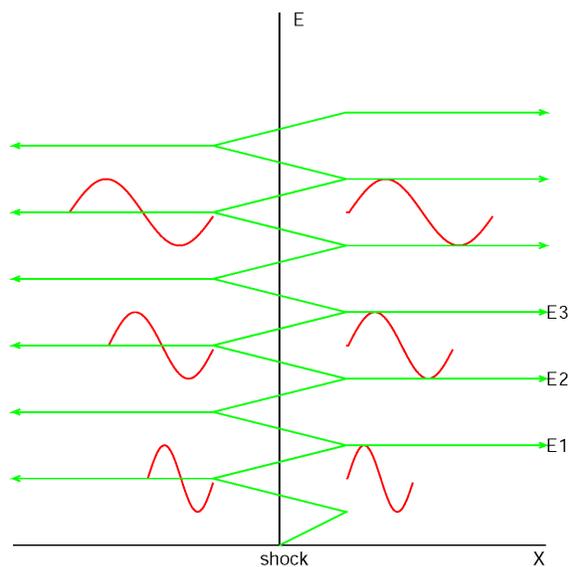

The first particles we see early in an event are those that have leaked out during the production of this shock-structure or ESP.  Shock acceleration begins at 2 or 3 solar radii (Reames 2009a, b; Cliver et al. 2004).  As the shock moves out in radius, plasma density begins to decrease and the flow of seed particles into the structure decreases; the local $B$ also decreases so existing waves resonate with particles of lower $P$.   Thus, the flow of protons upward in energy decreases and more of them flow out to renew the waves, so fewer arrive to renew waves at the top in Fig. 7.  Thus the highest energies leak away first and we see a trail of higher-energy escaped protons peaking well ahead of the shock arrival time in Figs. 2e and 3d.  GeV protons are only accelerated near the Sun in the largest events and may never be efficiently trapped, although much of the lower-energy ESP structure survives surprisingly well, even after 4 days and out to 2 AU or more (Fig. 3).  The big shock with its ESP event in Fig. 4 is still active at 4 AU after 13 days.  More-efficient quasi-perpendicular shocks aid in ESP survival.

Since streaming SEPs amplify waves, SEP events are involved in the business of creating ESP structures, but the SEP leakage during the process may exceed the trapping in the structure itself on some field lines.  The ESP peaks we see are actually a radial cross section of the shock structure which may also be quite extensive in latitude and longitude.  The ESP structure shows the radial thickness of the distribution of self-amplified waves, made visible by the particles they have trapped; it is a thickness that depends upon $\theta_{Bn}$ (e.g. Tylka et al. 2005) and thus distinguishes quasi-parallel and quasi-perpendicular regions of the accelerating shock wave (compare Figs. 2c and 2e).  In Fig. 3, any ESP event is obscured at *Helios 1* (Fig 3e), it emerges at *Helios 2* (Fig. 3f) and IMP 8 (Fig. 3c) and stands quite clear at *Voyager 2* in Fig. 3d.

The ESP structures often become overgrown in very large SEP events when the higher-energy particles that leak out early are intense enough and generate additional waves to trap lower energies before any begin to approach 1 AU.  These are the large events that reach the "streaming limit" and cause an early plateau region before shock arrival (Reames 1990, 1999, 2013; Reames and Ng 1998, 2010, 2014).  The streaming limit is easily explained:  ions streaming along the field amplify waves; the amplified waves increase ion scattering which reduces the streaming enough to reduce the wave amplification and produce equilibrium.  The higher-energy particles that arrive earliest generate waves that are sufficient to retard, flatten and roll over the low-energy intensities until the later increase near the shock itself (Reames and Ng 2010; Ng et al. 2003, 2012).  Only when the high-energy intensities are moderate, can the low-energy spectrum remain a power law.  Perhaps it can be said for these very large SEP events that the entire region, from the shock near the Sun out to the SEP onset at Earth and beyond, has become an immense wave-trapping ESP event as suggested in Fig. 8.  It is difficult to recognize these ESP events when we are actually inside them for as long as a day or so.  Extensive wave growth is an additional physical process that modifies the space-time profiles of the largest events.  The phases of SEP events including ESP events using proton-generated resonant waves have been studied theoretically by Lee (2005).

**Fig. 8** An overlay of intensity-time profiles of large SEP events shows similar streaming-limited plateau intensities with differing shock peaks for the six large SEP events listed (after Reames 1990). Does the extension of the hydromagnetic-wave amplification and SEP trapping from the shock peak into the whole plateau region in these large events warrant calling the entire events "ESP events" rather than just the peak at the shock? More trapping leads to higher energies at the shock.

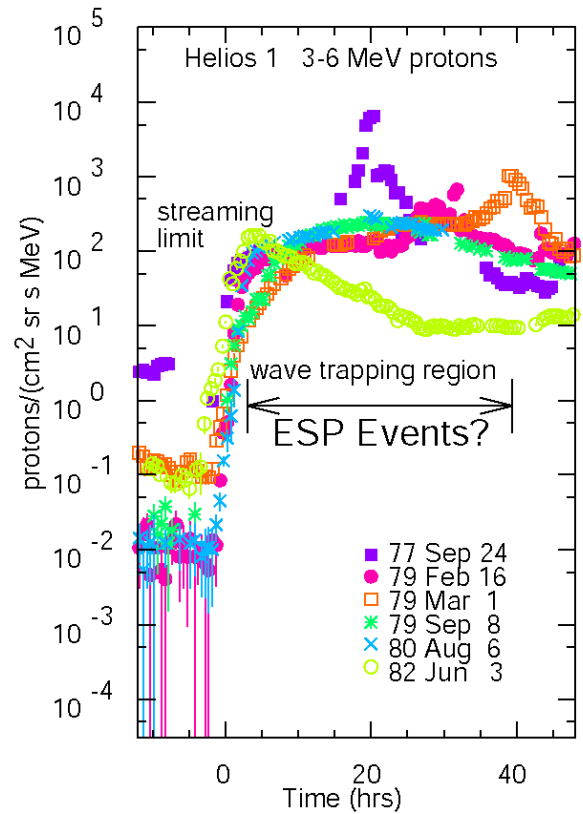

## 5 Discussion

The use of only peak intensities to characterize SEP events is a shortcut many of us once used, as in discussing the correlation of SEP intensities with CME speeds (e.g. Reames 2000a; Kahler 2001), for example. By assuming that a single point can represent a complex time profile, it became possible to easily study a large numbers of events. The correlations were subsequently improved by finding the CME speed at the base of the initial field line to the SEP measurements (e.g. Rouillard et al. 2012). Full three-dimensional fits to the CME spatial structures were then obtained (e.g. Kouloumvakos et al. 2019), but often the SEPs are still represented by a single point. In Sect. 3.4 we also found many attempts to fit SEP spatial distributions based upon three-point peak intensities, yet there were other events (e.g. Fig. 5) where the peak intensities had little or nothing to do with initial spatial distributions of shocks, but represented the time decay of a spatially-invariant reservoir. Furthermore, in Fig. 3d, the shock did not initially intercept the footpoint of the field line to Voyager 2 at all, but the shock and the peak proton intensities came upon it much later. There is more than one way a peak proton intensity can be produced. SEP events can be complex. Depending upon how "peak" is defined, some peak intensities can even be bounded by the streaming limit (e.g. the 3 June 1982 event in Fig. 8). For over half of the events shown in Fig. 2, below 50 MeV the peak is determined, not by the point on the shock at the footpoint of its field line, but by the local shock at the spacecraft. Peak intensities also depend upon the density and volume of seed particles and upon diffusion coefficients, not just shock speed. If there are maps of the spatial structure of CMEs and shock strength, would it not be better to construct models of the full time profiles of the SEPs, rather than just consider peak intensities?  Perhaps it is even possible to model the difference between the events in Fig. 2a and Fig. 2e or Fig. 2f.

It is time to extend the study of the relationship between CME-driven shock waves and SEPs beyond correlations with peak intensities to avoid "big flare syndrome" (Kahler 1982) which states that, statistically, many energetic phenomena become more intense along with larger flares, regardless of the detailed physics; everything correlated is not *caused* by flares.  Increased magnetic reconnection energy can spawn many increased phenomena not causally related to each other.  In small events everything tends to be small; in big events everything is big; of course there are correlations.  Can we, instead, relate the observed spatial distributions of the SEPs to specific spatial configurations of the CMEs and shocks, now uniquely observed (e.g. Kouloumvakos 2019), and thus better understand the detailed physics?  In this article we have tried to show that observed SEPs can sometimes be directly associated with locally observed shock waves in rather unique and inseparable ways; yet, other times they are associated with the reservoir.

If the full spatial configuration of the CME and shock is known but SEP modeling is too difficult, then there may be simple predictions that can be made: Which part of the shock is strongest and is most likely to dominate the SEPs, (1) the shock at the longitude of the spacecraft itself, or (2) the shock ~55° to the west at the footpoint of its field line? Of course we may also be able to predict (3) reservoir-dominated events like those of *Helios 2* and IMP 8 in Fig. 5.  For most SEP energies at both *Voyager* spacecraft in Fig. 4, the SEP peaks at the shock transit time equal or exceed the earlier peaks from their centrally-connected footpoints.  The second peak is larger than the first. For the 1 January 1978 event observed by IMP 8 (Fig. 3c) and *Helios 2* (Fig. 3f), the connection longitude and time of maximum intensity are a strong functions of energy that are explained by the radially diminishing shock strength.  Predicting the physical process and the dominant region of the shock goes a step beyond "big flare syndrome" (Kahler 1982) in saying how the source physics operates and not just the statistical correlation.

Other questions also emerge.  If the SEP longitude distribution in large events fits a Gaussian does the longitude dependence of the shock speed or shock strength also fit a Gaussian?  Three points always determine a Gaussian, i.e. a parabola in log space.  In general, how do the SEP intensities match the shock properties and how do they change in time?

We have assumed a single direct field line connecting an observer and an SEP source point at a shock.  However, it is well known that the random walk of magnetic field lines (Jokipii and Parker 1969) can map a confined location near the Sun to a distribution of points in space, or conversely.  This effect will spread the SEP distributions somewhat so that low SEP intensities may reach regions where they might not otherwise be expected.  This spreading depends upon the pre-event field, not upon the event itself, so it should not be an adjustable parameter that varies from event to event.  It is also true that interplanetary field configurations may be highly disturbed by previous CMEs (e.g. Gopalswamy et al. 2004) so that magnetic field connections are poorly approximated by simple Parker spirals in some cases. Multiple CME interactions can also modify peak SEP intensities.

Where should we look for naked ESP events?  Clearly they will come for spacecraft connected to the west flank of the shock where the Parker spiral allows shocks to propagate across field lines late in an event, after the early SEPs have streamed away. If the shock is strong, spacecraft farther from the Sun can benefit from the increased curvature of the field, like the *Voyager* spacecraft in Figs 3 and 4.  However, increasing distance may also allow shocks to weaken and fade before reaching the spacecraft, as seen by both *Helios 2* and IMP 8 in Fig. 5.  Here intensities rise but do not peak at the shock because the shock has faded, they finally peak at the reservoir that survives from earlier times.

Multiple encounters of a shock with a single field line will become common as the shocks reach radii where the field lines begin to completely encircle the Sun.  These multiple injections will become inevitable, even repeatable, for any shocks that remain sufficiently strong.  They will be difficult to find without the aid of a conveniently-located remote spacecraft like *Voyager*, and many could be lost as weak sunward-flowing increases in the clutter of multiple events at 1 AU.

While this article has focused on for the distributions of protons (and ions), non-relativistic electrons behave differently since they cannot resonate with Alfvèn waves as ions do.  Nevertheless, it has long been known (Wild et al. 1963) that type II and type III radio bursts involve streaming 10 – 100 keV electrons: type II emission at shock waves and type III emission from solar jets (Shimojo and Shibata, 2000).  The electrons that generate type II emissions may result from shock-drift acceleration at quasi-perpendicular regions of the shock and this emission may be confined near the shock because these electrons are more easily overtaken by the shock and swept downstream.  At 500 keV, an imaginative study by Cliver and Ling (2007) finds that electrons become well correlated with the shock-accelerated protons only in poorly-connected SEP events but not in well-connected events.  The well-connected regions must be dominated by the type-III-generating electrons from associated jets, not from the shock.  Other studies (e.g. Dresing et al. 2022) show good correlation of relativistic electrons with shock parameters. Unfortunately, however, these and earlier works do not consider the possibility of two spatially separated sources of electrons.

In this article we have discussed only protons, yet abundances of the other elements are extremely helpful in discerning the physics of SEP events.  Abundances of groups of elements have been studied up to the region of Pb (Reames 2000b; Mason et al. 2004; Reames and Ng 2004; Reames et al. 2014).  Abundances distinguish the "impulsive" and "gradual" SEP events and determine the relative importance of magnetic reconnection and shock acceleration (e.g. Reames 2020, 2022).  They measure the element abundances of the corona itself and their dependence upon the element's first ionization potential (FIP) and distinguish differences in the FIP dependence between SEPs and the solar wind (Mewaldt et al. 2002; Reames 2018a, b, 2020; Laming et al. 2019).  They measure SEP source temperatures and distinguish seed populations and acceleration mechanisms.  For a general discussion of SEP abundances see Reames (2021a) and references therein. Abundance measurements greatly contribute to our understanding of the physics of SEP events but are beyond the scope and focus of this article.

# 6 Conclusions

In an effort to characterize "typical" time profiles for SEP events from different solar longitudes, early observers overlooked some events, especially noteworthy are those dominated by intensity peaks at local shock waves that can survive autonomously as the historic "energetic storm particle" or ESP events. The neglect of these events has persisted. In hindsight, these local-shock-dominated events represent a fundamental opportunity to directly see underlying connections between SEPs and shock structure that are key to the physics of SEP acceleration. In contrast, the SEPs in reservoir-dominated events, often included, are related to regions of the shock that are much more remote.

As particles stream away from a shock, the protons amplify resonant hydromagnetic waves that trap subsequent particles, scattering them back and forth across the shock so they gain new energy and repeat this process. This is a fundamental process of diffusive shock acceleration that creates and maintains a population of SEPs quasi-trapped across an active shock by self-generated waves as the shock propagates outward from the Sun. The extent of this SEP structure in solar latitude and longitude is defined by the active shock itself. The radial thickness of the ESP event defines the spatial extent of the trapping wave field, made visible by the SEPs; it is also seen to depend upon $\theta_{Bn}$, thus distinguishing regions of quasi-parallel and more-efficient quasi-perpendicular shock acceleration. As this autonomous ESP structure moves outward its western flank can cross field lines of the Parker spiral, broadening the overall longitude span of the distribution to the west as seen by the westward march of the naked ESP event onto the new western field lines sampled by Voyager 2 in Fig. 3. These trapped SEPs along with those that have leaked away earlier, during the ESP formation and during its sustained motion, as well as those now trapped in a reservoir between the shock and the Sun, constitute the SEP distribution in space and time. Of course, shocks are strongest near the Sun, but far from the Sun we can still see shocks accelerate and trap protons of moderate 60 – 80 MeV energies, several AU from the Sun, for many days, perhaps weeks (e.g. Fig. 4), showing great residual power.

In events of small and moderate size, the SEPs that leak from the shock tend to stream away, leaving ESP structures bare after they cross field lines. When extremely fast shocks drive very large events, SEP intensities are high enough to create waves over a much larger field and the particle flows become bounded at the streaming limit, expanding the trapped ESP-like event over a huge volume of space and trapping more energetic particles at the shock, improving their rise to higher energies. We do not get to see naked ESP events here; those in extreme events are buried in trapped particles.

Peak proton intensities at different solar longitudes often come at different times, mixing spatial and temporal dependence and responding to different physical processes and different regions and properties of the shock. Peak intensities provide a poor characterization of SEP events. Better correlations with shock parameters may be possible by modeling and fitting the complex SEP event profiles using all known processes and following their time evolution. Perhaps we can relate the three-dimensional structure and time evolution of SEPs more completely to the three-dimensional structure and time evolution of observed CMEs and shock strength.

We have studied events where SEPs and the shocks that accelerate them are interwoven and inseparable and other regions where SEPs escape or are left behind. We have adapted the common theory of diffusive shock acceleration to help understand SEPs in terms of the persistent spatial structures that shocks and SEPs create together. We can best study the physics of SEPs and shocks when we can measure both together.

## Acknowledgement The author thanks Steve Kahler for helpful discussions and for his comments on the manuscript.

**Disclosure of Potential Conflicts of Interest** The author declares he has no conflicts of interest.